# Human-Machine Symbiosis, 50 Years On

Ian FOSTER
*Computation Institute, University of Chicago and Argonne National Laboratory
Mathematics and Computer Science Division, Argonne National Laboratory
Department of Computer Science, University of Chicago*

**Abstract**. Licklider advocated in 1960 the construction of computers capable of working symbiotically with humans to address problems not easily addressed by humans working alone. Since that time, many of the advances that he envisioned have been achieved, yet the time spent by human problem solvers in mundane activities remains large. I propose here four areas in which improved tools can further advance the goal of enhancing human intellect: services, provenance, knowledge communities, and automation of problem-solving protocols.

**Keywords**. Licklider, man-computer symbiosis, provenance, services

**Introduction**

In his classic 1960 paper, *Man-Computer Symbiosis* [22], L.C.R Licklider wrote of how monitoring his time spent at work led him to discover that:

> *About 85 per cent of my "thinking" time was spent getting into a position to think, to make a decision, to learn something I needed to know. Much more time went into finding or obtaining information than into digesting it. Hours went into the plotting of graphs, and other hours into instructing an assistant how to plot. When the graphs were finished, the relations were obvious at once, but the plotting had to be done in order to make them so. At one point, it was necessary to compare six experimental determinations of a function relating speech-intelligibility to speech-to-noise ratio. No two experimenters had used the same definition or measure of speech-to-noise ratio. Several hours of calculating were required to get the data into comparable form. When they were in comparable form, it took only a few seconds to determine what I needed to know.*
>
> *Throughout the period I examined, in short, my "thinking" time was devoted mainly to activities that were essentially clerical or mechanical: searching, calculating, plotting, transforming, determining the logical or dynamic consequences of a set of assumptions or hypotheses, preparing the way for a decision or an insight. Moreover, my choices of what to attempt and what not to attempt were determined to an embarrassingly great extent by considerations of clerical feasibility, not intellectual capability.*

These observations led him to advocate the use of computers to, in essence, "augment human intellect by freeing it from mundane tasks"—a goal that Doug Engelbart would soon also pursue, with great success [8].

Almost 50 years later, we have personal computers, immensely powerful software, huge online databases [18], and a ubiquitous Internet (another Licklider idea [21]). Our intellect has indeed been augmented: we can, for example, perform computations and data comparisons in seconds that might have taken Licklider hours, days, or years.

In other respects, however, the situation is less rosy. While we could probably process Licklider's six speech datasets in seconds rather than hours, we will probably still struggle with incompatible formats, and may well be dealing with six million or even six billion objects. Meanwhile, while the advent of the Web has dramatically increased access to data, it can still be exceedingly difficult to discover relevant data and to make sense of that data once it is located. And as we automate various aspects of the problem solving process, other activities emerge as the time-consuming "mechanical" steps. For example, in biology, DNA microarrays allows ten of thousands of measurements to be performed in the time that a researcher might have previously taken to perform a single measurement [35]. However, the amount of time per day that a researcher spends in "mechanical" labor may be no less: experiments must still be set up, data collected and stored, results analyzed. In other words, there are still just as many opportunities to automate the routine and mechanical.

This discussion emphasizes that as we near the $50^{th}$ anniversary of Licklider's paper, the need for man-computer symbiosis is no less urgent. However, we must demand far more from our computers than we did in 1960.

In the spirit of celebrating Licklider's legacy, I discuss here four related areas in which I believe significant progress can be made in further augmenting human intellect via the automation of the mundane and mechanical.

First, I examine how service-oriented architectures can make powerful information tools available over the network, for discovery and use by both people and programs. By permitting distribution of function, "service oriented science" (SOS) systems can both greatly reduce barriers to accessing existing intellectual tools—and permit (via the creation of networks of interacting services) the creation of new tools.

Second, I discuss how we can automate the documentation of data and computational results, so that users and programs alike can determine how much confidence to place in computational results. Such provenance mechanisms are an essential prerequisite to any serious attempt to realize SOS on a large scale.

Third, I point out how technology can facilitate the construction of effective communities, and thus increase the scale at which SOS techniques are applied and sustained.

Fourth and finally, I propose that the reach and impact of SOS, provenance, and community tools can be expanded by automating science protocols: extending the reach of automation to encompass not just simple computational tasks but also more complex procedures that may include experimental activities.

None of this material is new or rigorous. Nor is my review of the state of the art anything more than suggestive. Nevertheless, I hope that this presentation spurs some thoughts in my readers on where and how to advance the state of the art in scientific software and infrastructure.

## 1. Service Oriented Science

Emerging "digital observatories" provide online access to hundreds of terabytes of data in dozens of archives, via uniform interfaces [39]. These systems allow astronomers to pose and answer in seconds, and from their desk, questions that might previously have required years of observation in remote observatories. For example, astronomers can combine data from different archives to identify faint objects that are visible in the infrared spectrum but not the optical—so called brown dwarves [40].

In order to build such systems, astronomers have defined conventions for describing the contents of data archives and for the messages used to request and receive data. Thus, clients can discover and access data from different sources without writing custom code for each specific data source. These conventions address both low-level details of the format of the messages exchanged between clients and services and higher-level details concerning message contents. Web Services [5] specifications and software are widely used to address lower-level concerns; higher-level concerns tend to be addressed by more application-specific conventions, such as the VOTable specification [28].

Codified interfaces allow not only humans but also programs to access services. Indeed, it is arguably automated access by software programs that really makes such systems significant. In the time that a human takes to locate one useful piece of information, a program may access and integrate data from many sources and identify relationships that a human would never discover unaided. Thus, we can discover brown dwarves, integrate information automatically from genome and protein sequence databases to infer metabolic pathways [29], and search environmental data for extreme events.

Not only data but also programs that operate on data can be encapsulated as services, as can sensors, numerical simulations, and programs that perform other computational tasks. Networks of such services can be constructed that perform complex computational activities with little or no human intervention. Systems that are thus structured in terms of communicating services are called *service-oriented architectures.* I use the term *service-oriented science* (SOS) [12] to refer to scientific research assisted or performed by such distributed networks of interoperating services.

Many believe that SOS methods are vital for dealing with the rapidly growing volume of scientific data and the increasing complexity of scientific computing and research. In principle, SOS methods make it possible to decompose and distribute responsibility for complex tasks, so that many members of a community can participate in the construction of an eventual solution.

The successful realization of SOS is not simply a question of using Web Services or similar technologies to encapsulate data and software. We also need:

- **Resources** (data, software, sensors, etc.) that are viewed as valuable by multiple people, and reward systems that motivate people to construct and operate services that provide access to those resources. These "reward systems" can range from payment to peer approval and professional advancement.
- Supporting **software and services** that allow clients to discover services, determine whether services meet their needs, and make sense of results returned by services. Depending on context, these mechanisms can range from simple natural language descriptions of service capabilities and contents to sophisticated metadata, constructed according to agreed-upon ontologies, describing contents, provenance, and accuracy [36]. In many cases, authentication, authorization, and management mechanisms and policies are also required to control who can access services.

- The **hardware infrastructure**, operational support, and policies that allow services to be operated in a suitably convenient, reliable, secure, and performant manner, and that permit users to access services efficiently over local and wide area networks.
- A community of developers, operators, and users who have the technical expertise required to construct, operate, and use services. New approaches to **education and training** may be required to develop this community.

Note that success in each area depends on both technological and sociological issues. Indeed, the nontechnical issue of incentives may be the most important of all. A scientist may work long hours in the pursuit of not only knowledge but also tenure, fame, and/or fortune. The same time spent developing a service may not be so rewarded. We need to change incentives and enable specialization so that being a service developer is as honorable as being an experimentalist or theorist. Intellectual property issues must also be addressed so that people feel comfortable making data available freely. It is perhaps not surprising that astronomy has led the way in putting data online, given that its data has no known commercial value [39].

*1.1. Creating, Discovering, and Accessing Services*

For SOS to flourish, we need to kick start a virtuous cycle in which the following steps are performed repeatedly by many participants:

- Users discover interesting data and/or software services, and determine that they meet their purposes;
- They compose this service with others to create new capabilities; and
- They publish the resulting services for use by others (perhaps subject to access control).

We can thus catalyze the creation of distributed networks of services, each constructed by a different individual or group, and each providing some original content and/or value-added product.

The U.S. National Cancer Institute takes SOS seriously. Its caBIG project [34] (Figure 1) seeks to enable new approaches to cancer research and care by facilitating the sharing of data and software across the many cancer centers and related institutions. To this end, caBIG leaders have defined and implemented a comprehensive architecture that addresses every aspect of the service lifecycle, from authoring to publication, discovery, composition, and access. The resulting service oriented architecture builds on Web Services standards, vocabularies and ontologies developed within the medical community, and the Globus open source software [13].

Authoring is assisted by a tool called Introduce [17], which allows users to define stateful Globus-based services, specify deployment parameters, and specify access control policies for the new service. A complementary tool, the Remote Application Virtualization Environment (RAVE), builds on Introduce to allow for the wrapping of arbitrary applications as Web Services. Figure 2 shows the steps involved in Introduce-RAVE service creation and deployment:

1. Using the RAVE-enhanced Introduce, the application service is defined in terms of its executable, the form of its input and output messages, its access control policies, and other metadata.
2. The service code is generated and stored in a repository.
3. The service is also registered in a service registry, so that users can discover its existence.
4. When required (e.g., proactively, or in response to a user request), the service implementation is copied to an execution site …
5. … and deployed.

Once a service is deployed, users can then proceed to discover it and access it in the usual way:

6. A user or program can discover the service's existence …
7. … and invoke it via conventional Web Services mechanisms.

Of course, standard vocabularies are not always a prerequisite for automated analysis. To give one example, the GeneWays system mines the raw biological literature to identify experimentally derived relationships and to infer what credence to put in those relationships [33].

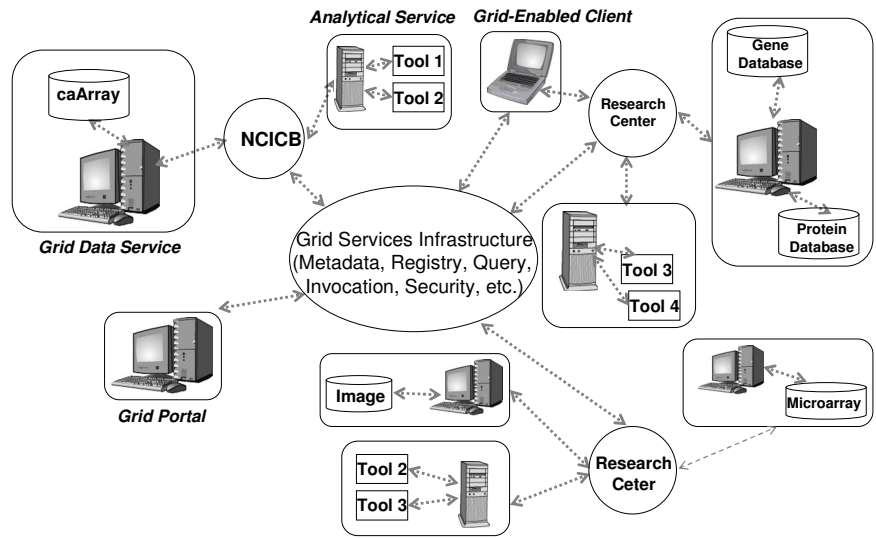

**Figure 1: A caBIG deployment, showing data and compute services, portals, NCI infrastructure, and other components**

*1.2. Hosting and Provisioning*

In order for this virtuous cycle to flourish, we must both minimize the costs of not only creating but also operating services and also make it possible to build services that can scale to meet application demands. Thus, we require efficient and convenient service hosting mechanisms. We take such mechanisms for granted when it comes to Web pages—few people run their own Web server nowadays—but they are still rare for services. These mechanisms should allow for the rapid and convenient deployment of new services, for the dynamic provisioning of services in response to changing demand, for access control, and for accounting and audit.

Service deployment mechanisms need (in one way or another) to acquire required resources at a hosting site, configure those resources appropriately, install and configure service code, and initiate the service. Globus Toolkit support for these functions illustrates some of the different ways in which they can be achieved:

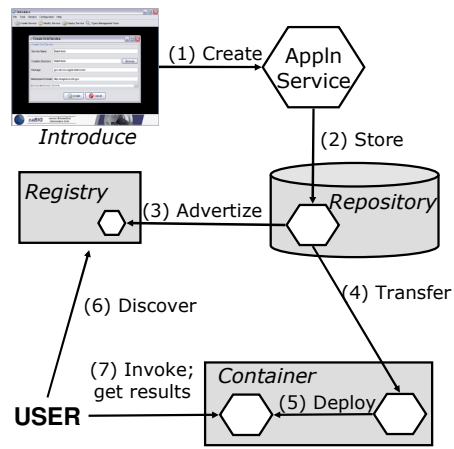

**Figure 2: Service creation and deployment steps**

- **Dynamic deployment** of Java Web Services into an existing container allows for the rapid and lightweight creation of new services [31]. However, this approach only works for Java Web Services, and the Apache Axis container that Globus uses does not provide for resource management among different services running in the same container.
- The **GRAM service** provides for the dynamic instantiation, and subsequent monitoring and management, of arbitrary executable programs, which may themselves then operate as services [9]. GRAM is widely used for service deployment, for example when "gliding in" Condor, Sun Grid Engine [41], or Falkon agents. However, GRAM only provides limited control over how the computational resource is configured.
- The **virtual workspace service** [19] provides for the dynamic deployment, and subsequent monitoring and management, of arbitrary virtual machine (VM) images. A VM instance provides a high degree of control over execution environment and resource allocations, but is a more heavyweight construct than a process.

Amazon's Extensible Computing Cloud (EC2) is one of several providers operating on-demand computing resources. Like the virtual workspace service, it provides a Web Services interface for virtual machine deployment and management; however, it provides only simple security mechanisms. A workspace service interface to EC2 makes it easy for service creators to deploy services onto EC2 resources.

Once a service is deployed, clients need to be able to monitor and manage its execution. They may also want to negotiate support for dynamic provisioning, i.e., for varying the resources allocated to a service in response to changing load. Services must often deal with data volumes, computational demands, and numbers of users beyond the capacity of a typical PC. Responding to a user request—or to the arrival of new data—can involve large amounts of computation. For example, the Argonne GNARE system searches periodically through DNA and protein databases for new and updated genomes and then computes and publishes derived values [37]. Analysis of a single bacterial genome of 4,000 sequences by three bioinformatics tools (BLAST, PFAM, and BLOCKS) requires 12,000 steps, each taking on the order of 30 seconds of run time. GNARE is able to perform these tasks in a timely fashion only because it has access to distributed resources provided by two U.S. national-scale infrastructures, TeraGrid [7] and Open Science Grid [38].

Dynamic provisioning becomes increasingly important as data analysis tasks are increasingly automated. For example, it is improbable that even a tiny fraction of the perhaps 500,000 biologists worldwide will decide to access Genbank, GNARE, or any other service at the same time. However, it is quite conceivable that 50,000 "agents" operating on their behalf would do so—and that each such agent would generate thousands of requests.

IBM's Oceano project [4] pioneered important ideas in dynamic provisioning, which is now becoming quasi-mainstream in certain commercial sectors. In my group, we are applying dynamic provisioning to both individual scientific applications and to scientific workloads with time varying resource demands. Falkon [32] monitors application load and then uses GRAM commands to acquire and release resources. By varying resource acquisition and release policies, we can tradeoff responsiveness to user requests and total resource consumption.

Finally, in a networked world, any useful service will become overloaded. Thus, we need to control who uses services and for what purposes. Particularly valuable services may become community resources requiring coordinated management. Grid architectures and software can play an important role [13]. We also need to be concerned with ensuring that SOS realize its promise of being a democratizing force, rather than increasing the gap between the "haves" and "have-nots."

## 2. Provenance

Progress in science depends on one researcher's ability to build on the results of another. SOS can make it far easier, from a mechanical perspective, for researchers to do just this, by using service invocations to perform data access, comparison, and analysis tasks that might previously have required manual literature searches, analyses, or even experiments. However, the results of these activities are only useful when published if other researchers can determine how much credence to put in the results on which they build, and in turn convince their peers that their results are credible.

One approach to this problem emphasizes reputation as the primary basis for evaluating and enforcing quality [44]. If each published result is associated with an author, then others can judge whether to trust a result based on their prior experience with results published by that author—and the author, being concerned with their reputation, will seek to maximize quality. This process is, in essence, that followed with print publications today, with the rigor of the reviewing process in a particular journal or conference also playing a role.

However, while reputation certainly has a role to play in trust, few researchers will be comfortable trusting a result on that sole basis. They will also want to see details on the method used to obtain a result. Such information can be used to determine whether a result can be trusted, can provide insights into when and where the result can be trusted, and can help guide new research.

Such documentation corresponds, in a broad sense, to the "methods" section in experimental papers, which should in principle provide enough information to allow a researcher to replicate an experiment. While that principle is perhaps honored more in the breach than in the observance, it is still a fundamental concept for science.

Increased use of computational techniques introduces new challenges to the documentation of experimental procedures (e.g., what version of software was used? what parameters were set?), but also offers the potential for significant improvements in "reproducibility." After all, while it may be impossible to capture the exact actions performed by an experimental scientist, the digital nature of computations means that it can be possible (in principle) to capture the exact sequence of computational steps performed during simulation or analysis.

These observations have motivated growing interest in methods for recording the provenance of computational results. Initial work focused on databases [6, 43], but interest has broadened to encompass arbitrary computations [11, 23]. A series of workshops [24] have led to the formulation of a provenance challenge [25], in which many groups have participated. Approaches explored include the use of functional scripting languages to express application tasks [45], file system instrumentation [27], and the use of a general-purpose provenance store [23].

### 3. Building Communities

Research occurs within communities, and the formation and operation of communities can be enabled by appropriate technology. Thus, Bill Wulf introduced in 1993 the concept of the collaboratory:

> *a center without walls, in which the nation's researchers can perform their research without regard to geographical location—interacting with colleagues, accessing instrumentation, sharing data and computational resources, and accessing information in digital libraries* [1].

Five years later, Carl Kesselman and I wrote that Grid technologies are concerned with:

> coordinated resource sharing and problem solving in dynamic, multi-institutional virtual organizations. *The sharing that we are concerned with is not primarily file exchange but rather direct access to computers, software, data, and other resources, as is required by a range of collaborative problem-solving and resource brokering strategies emerging in industry, science, and engineering. This sharing is, necessarily, highly controlled, with resource providers and consumers defining clearly and carefully just what is shared, who is allowed to share, and the conditions under which sharing occurs. A set of individuals and/or institutions defined by such sharing rules form what we call a* virtual organization *(VO).* [10]

These two characterizations capture important aspects of the technology required to enable collaboration within distributed communities, emphasizing in particular the need for shared infrastructure, on-demand access, and mechanisms for controlling community membership and privileges.

While great progress has been made in tools for forming and operating distributed scientific communities, many challenges remain. For example, mechanisms that work effectively for two or ten participants may not scale effectively to one thousand or one

million—not necessarily because implementations cannot handle the number of tasks, but because softer issues such as trust, shared vocabulary, and other implicit knowledge break down as communities extend beyond personal connections.

One approach to solving some scaling problems is to build infrastructures that allow clients to associate arbitrary metadata ("assertions") with data and services. Assuming that we can also determine whether such assertions can be trusted (perhaps on the basis of digital signatures, and/or yet other assertions), consumers can then make their own decisions concerning such properties as quality, provenance, and accuracy. Various popular systems demonstrate the advantages, costs, and pitfalls of different approaches to building such community knowledge bases: for example, the Wikipedia collaborative authoring system, the Flickr and Connotaea collaborative tagging systems [16], and game-based systems for improving tag quality [3].

### 4. Automating Protocols

Science is not simply a matter of analyzing data or running simulations. Depending on context, it can involve planning and conducting experiments, collecting and analyzing data, deriving models from data, performing many different simulations to explore the implications of models, inferring new hypotheses from data, and planning new experiments. As the complexity of each of these steps increases, each becomes a candidate for automation. Thus, we encounter several related concerns: identifying what to automate, determining how to automate, and documenting automated procedures so that they can understood, validated, and replicated.

In the natural sciences, a protocol is a:

> *predefined written procedural method in the design and implementation of experiments [that] should establish standards that can be adequately assessed by peer review and provide for successful replication of results by others in the field* [2].

These remarks were written in the context of procedures intended to be performed manually, albeit perhaps with the aid of automated equipment. However, they can also apply to procedures applied entirely by computers, in which case we may refer to an *automated protocol*.

Because automated protocols are performed by computers and without human intervention, they can operate far faster than manual protocols. Thus, it becomes increasingly important to document precisely what operations are performed. Arguably, the fact that operations are performed under computer control also makes it more feasible to describe the protocol's operation, although as Muggleton [26] notes, "there is a severe danger that increases in speed and volume of data generation in science could lead to decreases in comprehension of the results."

Two areas in which automation has already had a major impact are data collection and integration [15]. In astronomy, automated sky surveys collect many terabytes of digital data per year, enabling new approaches to astronomy, as discussed above. In biology, the cost of DNA sequencing has reduced from around $10 per base pair in 1990 to less than 1 cent per base pair today. Thus, it becomes possible to perform, for example, "genetic surveys" of many species, and integrate new data from different sources [29, 37].

Sky surveys and genome sequencing both involve a comprehensive survey of an entire object (the sky or genome). In other cases, decision procedures are required to guide data collection, as when searching for transient events in astronomy or when exploring combinatorial spaces, such as the result of one (set of) experiment(s) helps guide the selection of the next. In that case, automated protocols can involve not only data collection and analysis but also the decision procedures used to operate experimental apparatus. In one suggestive project, King et al. [20] describe a "robot scientist" that uses automated mechanisms to identify experiments that can discriminate among competing hypotheses. They report that their best algorithm can outperform humans in terms of number of experiments required to achieve a given accuracy of prediction. Such algorithms may become a standard part of the scientist's repertoire, and future papers may note that "we obtained these results using equipment X controlled by algorithm Y."

Technological improvements continue to reduce the cost and increase the speed of experimental apparatus. For example, microfluidic devices allow for cheaper and more easily automated laboratory experiments, by allowing the delivery of precise and minute quantities of experimental reagents. In an interesting twist, Prakash and Gershenfeld [30] describe how such apparatus can be controlled by embedded digital control, via what they call microfluidic bubble logic. Thus experimental protocols may extend to the configuration of multiple forms of digital and analog devices.

**5. Summary**

When Licklider expressed his vision of computer-human symbiosis, he was restating, in terms of the technology of his day, and with a particular focus on problem solving, Alfred Whitehead's observation that:

> *Civilization advances by extending the number of important operations which we can perform without thinking about them.* [42]

The computer has greatly expanded the number of operations that we can perform without thinking. However, as we have discussed in this paper, the number of operations that remain susceptible to automation remains large—indeed, is perhaps unbounded.

In seeking further opportunities for optimization of human problem solving, we need to take a system-level [14] or end-to-end view, in which we study and seek opportunities for optimization in every aspect of the problem solving process, not only by the individual researcher or within an individual laboratory, but also within and across communities. For example, we may determine that (as I have argued here) service oriented architectures can be used to distribute and thus accelerate the processes of publishing, discovering, and accessing relevant data and software; that the encoding of provenance information can facilitate the reuse of computational resources; that software support for building communities can promote the collaborative development of knowledge; and that the representation as data objects of the protocols used to perform experiments, analyze data, construct simulations, test simulation codes, and so forth, can raise the level at which the results of thinking ("cognitive artifacts") are reused. Many other opportunities can easily be identified.

In examining these issues, I have focused on the concerns of scientists and science. Scientists are certainly not alone in grappling with these issues. However, science is perhaps unique in the scope and scale of its problems and the subtlety of the questions that the methods discussed here can be used to answer. We may expect that methods developed for science can find application elsewhere, even as scientists look increasingly to computer science and information technology for tools that maximize the time that they spend thinking.

**Acknowledgements**

I am grateful to Lucio Grandinetti for the opportunity to present some of these ideas at the HPC 2006 conference in Cetraro, Italy, and to many colleagues for discussions on these topics, notably Charlie Catlett, Carl Kesselman, Rick Stevens, and Mike Wilde. This work was supported by the Mathematical, Information, and Computational Sciences Division of the Office of Advanced Scientific Computing Research, U.S. Department of Energy (DE-AC02-06CH11357).